\documentclass[runningheads]{llncs}

\let\lncsproof\proof \let\lncsendproof\endproof \let\lncsqed\qed
\let\proof\relax\let\endproof\relax

\usepackage{amsthm} %

\let\proof\lncsproof \let\endproof\lncsendproof \let\qed\lncsqed

\usepackage[T1]{fontenc}
\usepackage{graphicx}
\usepackage{algorithmic}
\usepackage{textcomp}
\usepackage[table]{xcolor}
\usepackage{hyperref}
\usepackage{subfig}
\usepackage{color}
\usepackage{url}
\usepackage{balance}
\usepackage{paralist}
\usepackage{fancyhdr}
\usepackage{verbatim}
\usepackage{booktabs}
\usepackage{float}
\usepackage[ruled]{algorithm2e}
\usepackage{multirow}
\usepackage{enumitem}
\usepackage{microtype}

\usepackage{amsmath,amssymb,amsfonts} %
\newtheorem*{myDef*}{Definition}

\newcommand{\ie}{i.\@\,e.\@\xspace}
\newcommand{\eg}{e.\@\,g.\@\xspace}
\newcommand{\etal}{et~al.\@\xspace}

\begin{document}
\title{Unveiling the Sentinels: Assessing AI Performance in Cybersecurity Peer Review}
\author{Liang Niu\inst{1} \and Nian Xue\inst{1} \and Christina P\"{o}pper\inst{2}} %
\institute{New York University, USA \and New York University Abu Dhabi, UAE}%
\maketitle              %
\begin{abstract}
Peer review is the method employed by the scientific community for evaluating research advancements. In the field of cybersecurity, the practice of double-blind peer review is the de-facto standard.
This paper touches on the holy grail of peer reviewing and aims to shed light on the performance of AI in reviewing for academic security conferences. Specifically, we investigate the predictability of reviewing outcomes by comparing the results obtained from human reviewers and machine-learning models.
To facilitate our study, we construct a comprehensive dataset by collecting thousands of papers from renowned computer science conferences and the arXiv preprint website. Based on the collected data, we evaluate the prediction capabilities of ChatGPT and a two-stage classification approach based on the Doc2Vec model with various classifiers.
Our experimental evaluation of review outcome prediction using the Doc2Vec-based approach performs significantly better than the ChatGPT and achieves an accuracy of over 90\%. 
While analyzing the experimental results, we identify the potential advantages and limitations of the tested ML models.
We explore areas within the paper-reviewing process that can benefit from automated support approaches, while also recognizing the irreplaceable role of human intellect in certain aspects that cannot be matched by state-of-the-art AI techniques.

\keywords{Peer Review  \and Cybersecurity \and AI \and ML \and ChatGPT.}
\end{abstract}
\section{Introduction}
\label{chap:introduction}

The scientific review and decision-finding process is 
highly relying on peer reviewers' judgment and agreement. 
The decisions finally rely on their judgment and discussion based on aspects like technical correctness, novelty, and coverage of experimental results~\cite{soneji2022sp} but also more subjective aspects like creativity, applicability, and scientific contribution. %
Even though biased%
, no better approach than human inspection is known for judging scientific progress. 
At the same time, advances in Artificial Intelligence (AI) and machine learning (ML)
raise the question, to which level AI models could act as a reviewer
in the scientific review process~\cite{srivastava2023day}.
Analyzing the difference between human review results and ML predictions can provide a way to uncover hidden aspects of the decision-making process.
The application of ML for the prediction of reviewing outcomes is interesting as an area as it challenges both the limit of AI and the logic of scientific publication in the first place. We expect that ML-based techniques can be used to predict a certain part of the review decision process, but not fully. %

We have set out to investigate this question for 
computer security and privacy since this research area has seen a tremendous increase in paper submissions in recent years, which has challenged the peer-reviewing process at first-tier security conferences\footnote{Commonly known as the ``Big-4'' top-tier conferences in computer security and privacy: ACM CCS, IEEE Security \& Privacy, ISOC NDSS, and USENIX Security.} and the reviewers' ability to provide timely and comprehensive reviews.
Organizers of those conferences have expanded the pools of reviewers, introduced journal-like paper revision opportunities and the submission of commented prior reviews, but the wealth of submissions remains challenging to handle.
We direct readers to Appendix~\ref{app:review_process} for a detailed introduction of the peer review process in ``Big-4''.
Recently, Soneji \etal \cite{soneji2022sp} investigated the peer-review process in the computer security field using qualitative research methods (interviews).
Our work is complementary as we conduct a quantitative investigation.

Related endeavors were pursued in other research areas of computer science.
The computer vision community presented methods for automatically deriving a measure of paper quality based on basic visual features of papers%
~\cite{von2010paper,huang2018deep}.
However, this line of research is solely based on the visual appearance of papers, ignoring the textual contents.
In Natural Language Processing (NLP), researchers proposed text-based methods to grade essays~\cite{larkey1998automatic}, answer mathematical questions~\cite{lan2015mathematical}, assess handwritten work~\cite{singh2017gradescope}, and evaluate papers ~\cite{yang2018automatic} (achieving an accuracy of below 68\%).
Due to the recent emergence of ChatGPT~\cite{ChatGPT}, there has been a surge in efforts to explore its application as an auxiliary academic reviewer or as a tool for assisting in paper reviewing~\cite{ChatReviewer2023,ChatPaper2023,srivastava2023day}.

In this paper, we investigate the performance of AI in the domain of cybersecurity academic paper reviewing.
Specifically, we develop a pipeline consisting of a Doc2Vec~\cite{le2014distributed} model for document embedding and ML classification models trained to predict acceptance or rejection based on these vector representations. 
We compare our proposed pipeline with ChatGPT. 
To construct a comprehensive dataset for training and testing, we gather a substantial collection of publicly accessible papers from leading computer science conferences, amounting to over 10,000 papers.
One significant challenge we face is obtaining a negative sample set for ML training. 
Since submitted papers are not publicly available until proceedings are published, we employ alternative approaches. 
We approximate a negative sample set by utilizing public archival versions of papers provided by authors and employ several heuristics in the selection process.

In short, the major contributions of this paper are:
\begin{compactenum}
  \item We build a large dataset with over 14,000 papers. It consists of over 10,000 accepted conference papers and over 4,000 preprint papers collected from the arXiv preprint website.
  \item We train ML models to predict whether a paper is to be accepted by top-tier conferences in computer security and privacy and show their quantitative predictive results.
  Our experiments show that our best models can achieve approx.~91\% accuracy in predicting the reviewing decisions for security papers, significantly outperforming ChatGPT.
 \item We conduct further experiments to explore the capability of our method in dealing with abstracts and novel papers. 
 Subsequently, we analyze the experimental results, present the insights gained, and discuss the implications of both ChatGPT and the proposed pipeline in relation to the current peer-review system.

\end{compactenum}

\section{Related Work and Background}%
\label{chap:background}

\subsection{Related Work}
Several approaches have been proposed for similar tasks~\cite{lin2023automated}, broadly classified into two categories: vision-based and text-based approaches.

\noindent \textbf{Vision-based Approaches. }
Von Bearnensquash~\cite{von2010paper} proposed a method based on AdaBoost to classify an academic paper using the paper's appearance (\ie, paper gestalt). The author first turned papers into pictures using pdf-to-image conversion tools and then trained a classifier using the image features extracted from these pictures. 
Later on, Huang~\cite{huang2018deep} extended this idea by leveraging deep-learning models.
The author
also built a generative model that could generate paper gestalts that would be accepted by the classifier. 
The author concludes figures and tables are crucial factors for predicting the decision. 

However, these trained models are not suitable for predicting reviewing results of security papers since security papers differ from computer-vision papers in general. %
Essentially, vision-based methods only consider paper layout %
and neglect paper contents%
. When papers are being reviewed, the criteria are focused on the contents (\eg, writing quality, novelty, contribution, etc.).  That is why our method considers text content rather than paper layout as a feature.

\noindent \textbf{Text-based Approaches. }%
Taghipour \etal~\cite{taghipour2016neural} proposed
neural network models including CNN and LSTM~\cite{hochreiter1997long} for the task of automated essay scoring. The authors compared neural networks with different settings in their experiments. 
Alikaniotis \etal~\cite{alikaniotis2016automatic} also leveraged neural networks in their paper. The authors proposed an augmented C\&W model~\cite{collobert2008unified} with LSTM to score essays on a Kaggle dataset. Even though their methods are relevant to ours, their aim is focused on scoring educational essays rather than scientific papers, which fundamentally differs from our goal.
We consider a 2018 paper from Yang \etal~\cite{yang2018automatic} to be the most relevant research to our goal. They first proposed a new task called automatic academic paper rating. The authors built a new dataset using papers collected from arXiv and proposed a modularized hierarchical CNN as a classifier. Their method is claimed by the authors to be the state-of-the-art method to rate academic papers. However, they do not focus on papers on computer security and insights into the peer-review paradigm.
Another interesting research was conducted by Bartoli \etal~\cite{bartoli2016your} %
who proposed a model for the automatic generation of scientific paper review comments. Their goal is to generate reviews for scientific papers that could deceive people. They use traditional methods rather than neural networks to avoid the requirement of a large amount of training data.

Our work differs from these related works as follows: (1) Our model is a classification model which predicts the decision of an academic paper. (2) We focus on computer security conference papers. (3) Our discussion is focused on the peer-review process. (4) We provide comparison with ChatGPT.

Besides the aforementioned technical works, a recent IEEE S\&P paper by Soneji \etal~\cite{soneji2022sp} is a great inspiration. They conducted a qualitative study of 21 reviewers and chairs to understand the peer-review process in computer security. Over half of the participants shared negative sentiments toward the current review system. Their findings motivate us to explore the predictability of peer-review outcomes for security papers.

\subsection{Background}
\noindent \textbf{Doc2Vec. }
Doc2Vec~\cite{le2014distributed}, as its name suggests, is a method to represent a document by a vector. It is a method developed on the basis of Word2Vec. 
Doc2Vec is also an unsupervised learning method and it could benefit from a large dataset. We leveraged this feature of Doc2Vec and trained our model on a large set of academic papers to enable Doc2Vec to generate high-quality document embedding for the ML-based reviewing process.

\noindent \textbf{ChatGPT. }
ChatGPT~\cite{ChatGPT}, developed by OpenAI, is an advanced chatbot that utilizes a large language model (LLM) and state-of-the-art language modeling techniques to generate responses that closely resemble human-like conversations.
Through extensive training on diverse textual data, ChatGPT demonstrates an impressive ability to comprehend and produce coherent and contextually appropriate replies.
Notably, recent research suggests that GPT model represents an early version of an artificial general intelligence (AGI) system, although it remains incomplete~\cite{bubeck2023sparks}.
The integration and utilization of ChatGPT in academic settings have sparked discussions~\cite{naturecouldai,naturewhatchatgpt}.

\section{Data Collection}
\label{chap:dataacquisition}
To train and test the paper classification models, a dataset with a large number of scientific papers is needed.
Since there exists no dataset which can be utilized directly, we first need to create such a dataset.

Given the unsupervised nature of Doc2Vec's learning approach~(See Section~\ref{sec:pipeline}), its performance improves with larger training data.
However, the limited number of security conference papers impedes the training of Doc2Vec. To overcome this limitation, we gather accepted papers from prominent computer science conferences outside the realm of security.
These conferences cover a wide range of topics and employ shared vocabularies, making them suitable for pre-training the Doc2Vec model.
To generate high-quality embedding representations for security papers, we adopt a transfer-learning approach: 
Our Doc2Vec model undergoes initial training on general computer science papers, followed by fine-tuning on computer security papers.

To create the datasets, we collected published papers from first-tier computer science conferences encompassing domains such as computer vision, networking, and security, among others. 
We also incorporate papers sourced from the arXiv preprint website into our training data as negative (reject) samples; the details are explained in the following subsection.

\subsection{Dataset Composition}
Our dataset contains two subsets: Proceedings and Preprints. An overview of the these two datasets is provided in Table~\ref{tab:datasetdiff}.

\begin{table*}[bt]
		\centering
		\caption{The two subsets of papers used in our approach.
  }
		\resizebox{\textwidth}{!}{
            \begin{tabular}{@{}lccccr@{}}
				\toprule
				Subset & Component & Label & Venue & Amount & Usage \\ 
                    \midrule
				1: Proceedings & Accepted  papers   & Accept & \begin{tabular}[c]{@{}c@{}}Top-tier computer science \\ conferences\end{tabular} & 10519 & \begin{tabular}[r] {@{}c@{}}Doc2Vec + Classification (Big-4\\ papers); only Doc2Vec (others)\end{tabular} \\ \midrule
				2: Preprints      & 
				arXiv  preprints & Reject & \begin{tabular}[c]{@{}c@{}}Drafts from security  conferences\\ and other security preprints\end{tabular}              & 4220         & Doc2Vec + Classification                                                                                     \\ \bottomrule
			\end{tabular}
		}
	\label{tab:datasetdiff}
 \vspace{-10pt}
\end{table*}

The \textbf{proceedings subset} consists of over 10,000 \emph{published} papers sourced from top computer conferences, encompassing both security-specific venues and broader computer science domains. Detailed statistical information regarding the conferences, their venues, and sample sizes can be found in Table~\ref{tab:cpd} (Appendix).

The \textbf{preprints subset} comprises papers obtained from the arXiv preprint website. This subset plays a crucial role in the second stage of training, where our proposed models aim to determine the acceptance or rejection of papers by the ``Big-4'' conferences. We opted to utilize the arXiv dataset as negative (rejected) paper samples, employing well-defined selection rules (outlined below). This approach serves as an approximation of real review outcomes from actual security conferences, as acquiring such data directly presents challenges both logistically and ethically: First, accessing review processes is difficult due to the closed and confidential nature of the review process. Second, it would be inappropriate to collect papers and review outcomes without approval from the authors and reviewers at the time of submission.

Thus, we focus our attention on public sources of data, and after careful consideration, we have identified arXiv as a viable and rational choice.
Widely embraced by the computer science community, arXiv offers a substantial volume of training samples.
Furthermore, its research-friendly policy allows us to utilize the data without ethical concerns. We acknowledge that this approach is not without limitations. Extracting papers from arXiv, which are likely to have been submitted to and rejected by the "Big-4" conferences, is subject to errors and serves only as an approximation of actual review outcomes. Nonetheless, we contend that our meticulous selection process, guided by heuristic rules, sufficiently reflects reality.
In particular, we use published papers as positive samples, while negative samples are selected from the arXiv preprint papers based on specific criteria outlined below. The selection process for negative samples focuses on papers labeled with \texttt{cs.CR} on arXiv (indicating they belong to the security domain) and employs the three following heuristic rules:
\begin{itemize}[leftmargin=0cm,itemindent=.5cm,labelwidth=\itemindent,labelsep=0cm,align=left,topsep=0pt]
  \item \textbf{Rule 1}: For arXiv preprint papers that ultimately appear in the ``Big-4'' conferences, if one paper has multiple versions on arXiv and its first version is at least one year older than its last version, then we consider the first version paper as the negative sample. 
  This rule is based on the reasoning that, if a paper got accepted by the ``Big-4'', and it has evolved for at least a year before the acceptance, then we believe its first version is already good but just not of enough quality to be accepted by the ``Big-4'', which makes it suitable to be used as a negative sample. 
  Considering the improvement of content also takes time, we choose one year as a crude estimation of the heuristic threshold based on our experience in refining the papers.
  
  \item \textbf{Rule 2}: For arXiv preprint papers that have finally appeared on lower ranked security conferences\footnote{The lower ranked security conferences we picked for Rule~2 are: ESORICS, RAID, ACSAC, DSN, CSFW, ASIACRYPT, SecureComm, AsiaCCS, ACNS, SAC, ACISP, ICICS, ICISC, ISC, FSE, WiSec, SEC, SACMAT, CT-RSA, DIMVA, and ISPEC.}, we consider their first arXiv version as the negative sample. 
  The reasoning here is that, if a paper got accepted by lower ranked security conferences, then it is possible they were first submitted to the ``Big-4'', got rejected and then submitted to other conferences. Even if this is not the case, we believe these papers are of good quality, but just not enough to get accepted by the top venues in the field. 
  
  \item \textbf{Rule 3}: If an arXiv preprint paper does not get published at all and it was created prior to 2018, then we consider it as a negative sample. This rule serves as a supplement to ensure the generation of a balanced training set when the first two rules do not yield a sufficient number of samples. The rationale behind this rule is that preprint-only security papers are likely submissions to the "Big-4" conferences or other reputable security conferences that did not receive acceptance. We limit the inclusion of preprint papers created before 2018, as more recent papers may still be undergoing the review process.
\end{itemize}

We obtained a total of 43, 423, and 3,754 papers based on the three rules, respectively.
It is worth noting that not all of these negative samples were utilized for training.
With the proposed three rules, we identified more negative samples in the arXiv dataset than the positive samples from ``Big-4'' conferences. 
To balance the ratio of positive and negative samples, some papers selected from Rule 3 are randomly dropped.
Since there are $3984$ positive samples, the number of negative samples was adjusted accordingly.
The final numbers of negative samples used for training, based on the three rules, are 43, 423, and 3,518, respectively.

\subsection{Data Acquisition}
All papers in the proceedings subset we collected can be downloaded from public websites.
We developed automated crawlers to retrieve a subset of accepted papers from various conference websites, supplemented by manual downloads for the remaining papers.

For the preprints subset, we first fetch the arXiv metadata from Kaggle\footnote{\url{https://www.kaggle.com/Cornell-University/arxiv}}, and then use a crawler to download all the cs.CR papers from \url{export.arxiv.org}. 
We further use the DBLP~\footnote{\url{https://dblp.org/}} search function to check the venue information for all the downloaded papers so that we could select negative samples based on the heuristic rules.
The DBLP team has a rigorous process of quality checking all new additions to the database~\cite{dblpFAQrigor}. 
The search function of DBLP is powered by CompleteSearch~\cite{bast2007completesearch}.
We conducted tests to assess the efficacy of the DBLP search function. The retrieval results were found to be accurate, even in cases where a few words were missing from the paper's title. It is important to note that significant changes in a paper's title may limit the effectiveness of the DBLP search function. However, we contend that the core keywords within the title generally remain consistent, reinforcing the reliability of the search function even in instances where the title undergoes substantial modifications.

\section{Methodology}
\label{chap:model}

\begin{figure}[tb]
    \centering
    \includegraphics[width=.8\linewidth]{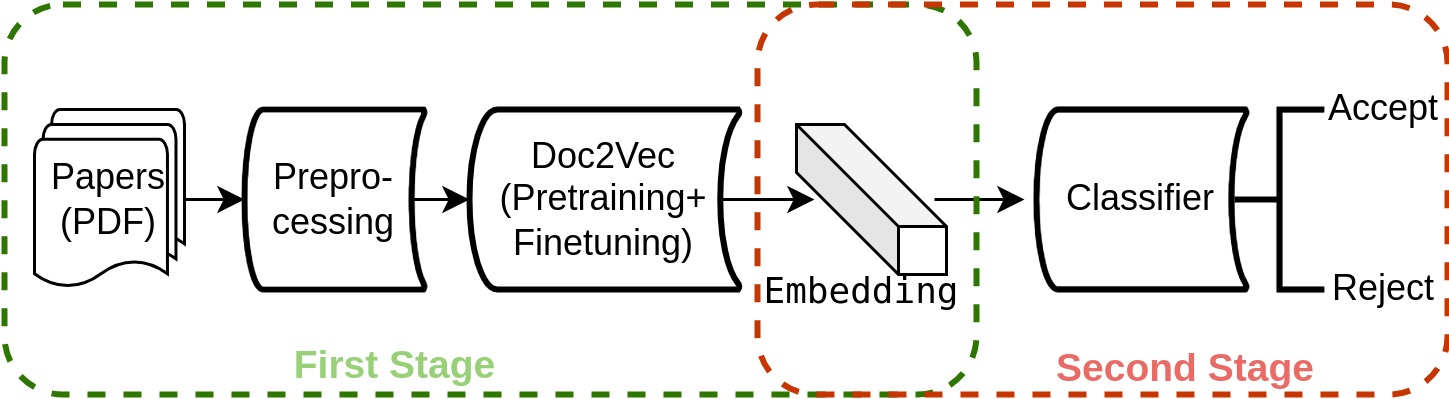}
    \caption{The pipeline of Paper Classification Model. The left-hand side is the first stage, where a Doc2Vec model turns papers into document embeddings. The right-hand side is the second stage, where classification models predict accept or not based on the given embeddings.}
    \label{fig:pipeline}
    \vspace{-10pt}
\end{figure}

We first present our terminology (Sec.~\ref{sec:terminology}) before we elucidate the fundamental principles of vector embeddings as basic problem definition (Sec.~\ref{sec:pipeline}). We then delve into the intricacies of our paper classification process (Sec.~\ref{sec:preprocessing}), Figure~\ref{fig:pipeline}), and outline our methodology for conducting the ChatGPT experiment (Sec.~\ref{sec:chatgpt}).

\subsection{Terminology}
\label{sec:terminology}

The reviewing process can be viewed as a text classification problem, and its formal definition is given here. We gathered papers from top conferences and let them be a set $\mathbb{D}= \left\{d_1,\dots,d_N\right\}$, where $N=|\mathbb{D}|$ is the number of papers. Each paper consists of $M$ words and is represented as $d_i= \left\{w_1,\dots,w_M\right\}$, where $M$ is the number of words in the paper. Then the raw papers are mapped to a high-dimensional vector space by the Doc2Vec model, after which each paper $d_i$ is represented as $\vec{d_i}$.

\begin{myDef*}
    The binary classifier, denoted $f(\cdot)$, is the function defined by 
    \begin{equation}
        f(\vec{d_i}) = \left\{\begin{matrix}
             1,& if\ \eta(\vec{d_i})>1/2\\ 
             0,& otherwise,
        \end{matrix}\right.
    \end{equation}
\end{myDef*}
where $\eta(\cdot)$ is an ML classification algorithm, giving out a prediction probability score between $[0, 1]$ based on the input vector. ``1'' means the classifier predicts the acceptance of the corresponding paper, while ``0'' indicates rejection.

\subsection{Problem Definition: From Papers to Vector Embeddings}
\label{sec:pipeline}

Here we briefly introduce how to use Doc2Vec to transform an original paper into a high-dimensional vector.

In the Doc2Vec framework, every raw paper is projected to a unique document vector~\cite{le2014distributed}, represented by a column in the matrix 
\[
\mathbf{D} =
\begin{bmatrix}
     \vec{d_1} & \dots &\vec{d_N} 
\end{bmatrix}^T,
\]
and every word in the paper is also mapped to a
unique vector during training, represented by a column in a matrix $\mathbf{W}$.
Given a paper $d_i$ and a sequence of words in the paper $w_1,\dots,w_M$, the objective function of the Doc2Vec model is to maximize the average log probability function
\[
J(\theta)=\operatorname*{argmax}_\theta \frac{1}{M} \sum_{t=k+1}^{M-k} \log p(w_{t} |\vec{d_i}, w_{t-k}, \cdots, w_{t+k}; \theta),
\eqno{(2)}
\]
where $k$ is the size of the training context and $\theta$ denotes the model parameters. The document vector $\vec{d_i}$ here represents the missing information from the current context and can act as a memory of the
topic of the document. Given the linear nature of the text, the contexts are fixed-length and sampled using a sliding window over the consecutive words.
For machine learning problems, we generally like to minimize the value of a cost function, so a negative sign will be added to the right side of $J(\theta)$ to make it a cost function $J^{\prime}(\theta)$ in training.

The context probability can be obtained by a multiclass classifier, \eg, softmax function. 
Accordingly, we have
\[
p(w_{t} |\vec{d_i}, w_{t-k}, \cdots, w_{t+k}) = \frac{exp(y_{w_t})}{\sum_j exp(y_j)},
\eqno{(3)}
\]
in which all the $y_i$ are the logits output from a neural network used for training.

Since it is a problem finding the minimum of the cost function $J^{\prime}(\theta)$, it is natural to use the gradient descent optimization algorithm.
The gradient is obtained via backpropagation. And then we use gradient descent to train the document vectors and word vectors and update the parameters in the model. 

After the training stage, we employ the model to perform an inference stage to calculate the document vector for a new paper. The outcome is then  directly fed into machine-learning classifiers, \eg, Na\"{\i}ve Bayes, Logistic Regression, support vector machines, or KNN for prediction.
Finally, we use these classifiers to predict whether the paper will be accepted or not.

\subsection{Method: From Preprocessing to Classification}
\label{sec:preprocessing}

Our proposed methodology comprises three primary stages (see Figure~\ref{fig:pipeline}):
\begin{compactenum}
 \item \textbf{Preprocessing:} First, PDF files are converted into textual data, after which anonymization is performed by removing author names and affiliations. Subsequently, the corpus is normalized.

 \item \textbf{Obtaining document embedding:} Then, we turn the normalized corpus into document embeddings (latent space representations) using a trained Doc2Vec model. 

 \item \textbf{Classification:} Finally, we use classification algorithms to classify the embeddings. Different algorithms are tested for comparison. 
\end{compactenum}

\noindent\textbf{Preprocessing of Collected Data: }
The collected papers are  processed into the proper form to ensure their compatibility with subsequent modeling steps.

\begin{compactitem}
 \item \emph{PDF to Text.}
All collected PDF papers require conversion to plain text for compatibility with NLP models such as Doc2Vec.
The tool used for such a task is the open source software PDFMiner~\cite{pdfminer}. 
 \item \emph{Text Normalization.} \ 
Following the conversion of PDF to plain text, we normalize the text to enhance efficiency during Doc2Vec training. The normalization techniques that we use in the preprocessing include: Contraction Expansion, Lemmatization and Stemming, Stop  Word  Removal, etc. 
In addition to conventional text normalization techniques, we employ task-specific methods to process and normalize the corpus. This involves removing extraneous content such as author information, bibliography, and publication-related details from the documents.
These techniques ensure that our pipeline is double-blinded just like the double-blind peer-review process, which is different from previous NLP-based researches that often take account of the authors~\cite{yang2018automatic}.
\end{compactitem}

\noindent\textbf{Document Embedding: }
Different methods can be used to generate document embeddings (Section~\ref{chap:background}).
Our choice is to use Doc2Vec for the proposed automatic paper reviewing pipeline because it is simple yet powerful, unsupervised, and able to get better as the training set becomes larger. 
To enhance the embedding generation capability of the Doc2Vec model for computer security papers, a two-phase training approach is employed. In the initial pretraining phase, the model is trained on a comprehensive dataset consisting of both security and non-security computer science papers, utilizing papers from both subsets. Subsequently, in the second phase, the model is fine-tuned exclusively on security papers to enhance its sensitivity to the security domain.

\noindent\textbf{Classification: }
In our experiments, 14 widely used ML classification algorithms are tested. %
Our negative sample in the preprints subset consists of more than $4,000$ papers. However, the number of ``Big-4''-published papers we could collect in the proceedings subsets is not enough,
which creates an imbalanced ratio of positive and negative samples.
This may significantly affect classification results, because the algorithms all assume a 50\% to 50\% ratio of positive and negative samples. If they are trained with such an imbalanced dataset, the classifiers will learn to lean toward rejection over acceptance.
To address the imbalanced data issue, we first tried SMOTE (Synthetic Minority Over-sampling Technique)~\cite{chawla2002smote} to resample the data, however, the results were not promising. Hence, we decided to abandon a randomly selected part of the negative samples selected from heuristic rule 3, to make the number of negative samples equal to the number of positive samples. 
Finally, we work with $3,984$ positive samples (all are published ``Big-4'' papers) and $3,984$ negative samples (arXiv preprints selected by heuristic rules).

\subsection{ChatGPT-based method}
\label{sec:chatgpt}
We limit the testing data for ChatGPT experiment to papers published after Sept. 2021, aligning with the training data and knowledge cutoff date of ChatGPT 3.5. Since ChatGPT possesses information regarding review outcomes for papers preceding this date, we focus on utilizing data from after September 2021 to enable ChatGPT to predict unseen papers rather than reiterating its existing knowledge. Specifically, 994 published papers from CCS '21, CCS '22, NDSS '22, NDSS '23, S\&P '22, USENIX '22 are used as positive samples and 994 papers from arXiv dataset are used as negative samples for testing. 
The papers are first converted into plain text and then split into chunks due to the token limitation of ChatGPT. 
For the prompt used in the ChatGPT experiment, we direct readers to Appendix~\ref{app:prompt}.

\section{Experimental Evaluation}
\label{chap:experiment}
We conduct experiments to report the predictive results of the proposed method and report the results on ChatGPT.

\subsection{Experimental Results}
As a baseline for comparison, we assume that there is an oracle that randomly determines if a paper will be accepted or rejected without bias. Therefore, the accuracy that the oracle gives a correct prediction is $50\%$ all the time. We compare  various ML models to this Random Guess baseline.
We tested 14 different classification algorithms, as well as two model ensemble methods, Voting and Stacking. The final classification results regarding each algorithm are shown in Table~\ref{tab:result}. Among all the classifiers, the average accuracy on testing data is $85.56\%$ and the median accuracy is $88.14\%$. The average F1 score is $0.8492$ and the median F1 score is $0.8809$, respectively. Moreover, most classification algorithms could finish training and testing in half a minute. 
In fact, the inference time of all these models could almost be neglected, because most of the time is spent on training. 

\subsection{Results of Model Ensemble}

Model ensemble is the method that uses a set of ML models to obtain better predictive performance than any of a single learning model alone \cite{polikar2006ensemble,rokach2010ensemble}. By doing so, the final results can ``learn from each other's strength'', integrate the learning capabilities of each model, and improve the generalization ability of the model. Thus, we select five models (\ie, Logistic Regression,  RBF SVM, Gaussian Process, AdaBoost and LDA) out of the 14 tested models and utilize ``Voting''\cite{witten2002data} and ``Stacking''\cite{wolpert1992stacked} strategies for model ensemble and then predict the results. The selection of models follows the principle of diversity and performance to ensure high robustness and high accuracy of the ensemble model. The final estimator we used in ``Stacking'' strategy is a decision tree classifier.

\begin{table}[tb]
\centering
\vspace{-1.0em}
\caption{\textbf{Experimental results.} Time refers to training time + testing time.}
\resizebox{0.75\columnwidth}{!}{ 
\begin{tabular}{@{}lcccr@{}}
    \toprule[1.5pt]
    \textbf{Algorithm}      & \textbf{Accuracy} & \textbf{F1 Score} & \textbf{AUC} & \textbf{Time (second)} \\ \midrule
    Random Guess  & 0.5 & 0.5  & 0.5    & -  \\
    \midrule
    Voting on Abstract   & 8306  & 0.8303 & -    & 641.15    \\ 
    \midrule
    Voting Classifier    & 0.910 & 0.910 & -     & 962.75    \\
    Stacking Classifier  & 0.851 & 0.851 & 0.851 & 3643.98    \\
    \midrule[0.5pt]
    Logistic Regression  & 0.912 & 0.912 & 0.959 & 0.25    \\
    Linear SVM           & 0.918 & 0.918 & -     & 9.08    \\
    RBF SVM              & 0.876 & 0.876 & -     & 13.98    \\
    Poly SVM             & 0.899 & 0.899 & -     & 23.72    \\
    Gaussian Process     & 0.910 & 0.910 & 0.962 & 877.59    \\
    Decision Tree        & 0.720 & 0.720 & 0.719 & 2.41    \\
    Random Forest        & 0.880 & 0.878 & 0.946 & 12.50    \\
    Boosted Tree         & 0.878 & 0.877 & 0.939 & 113.39    \\
    MLP                  & 0.903 & 0.903 & 0.957 & 14.52    \\
    AdaBoost             & 0.883 & 0.883 & 0.954 & 0.85    \\
    Gaussian Naive Bayes & 0.830 & 0.827 & 0.882 & 0.08    \\
    KNN                  & 0.565 & 0.469 & 0.825 & 23.64    \\
    LDA                  & 0.903 & 0.903 & 0.960 & 0.31    \\
    QDA                  & 0.853 & 0.850 & 0.918 & 0.20    \\
    \midrule
    Average              & 0.8556 & 0.8492 & 0.9062 & 356.20 \\
    Median               & 0.8814 & 0.8809 & 0.9428 & 13.24  \\
    \bottomrule[1.5pt]
\end{tabular}
}
\vspace{-2.0em}
\label{tab:result}
\end{table}

The model ensemble results are shown in the ``Voting Classifier'' and ``Stacking Classifier''  rows of Table~\ref{tab:result}. According to the results, we can find that both ensemble models' accuracy and F1 score are higher than $0.85$. Especially, the ``Voting'' strategy achieves $0.91$, which is similar to the best results in all the classification algorithms we test. Theoretically, an ensemble model is usually more robust than a single classifier. So in the following discussions, we will use the ``Voting'' strategy classifier as the representation of the experimental result.
Generally, the obtained accuracy and F1-scores per model are rather close to each other, indicating that the false negative and false positive rates are similar.

\subsection{Using Abstracts for the Prediction}
One interesting idea worth to further explore is to see the accuracy of merely using the paper's abstracts for prediction. 
To explore this question, we extract the Abstract section from raw papers and then use a Doc2Vec model trained for paper abstracts to convert them to high-dimensional embeddings. Specifically, we look for the text between ``Abstract'' section and ``Introduction'' section. And if these two sections are not able to be located, then we use the first 2,000 characters in the documents to represent abstract as a remedy. The extracted abstract is normalized and then used for training and testing.

The experimental result on the abstract-only prediction is shown in Table~\ref{tab:result} ``Voting on Abstract'' row. 
The accuracy and F1 scores we get using the ``Voting'' classifier are $83.06\%$ and $0.8303$, respectively. Compared to whole paper prediction, the performance degrades when we predict the result merely relying on the abstract, which is consistent with our intuition. In general, an abstract includes a paper's core idea, methodology, experimental results (\eg, whether it achieves the SOTA performance), etc. from a high level. 
That is to say, the abstract is a good indicator of the corresponding paper's applicability, and scientific contribution which are usually used as the criteria to determine whether a paper should be accepted or rejected. 
Although there is performance loss, it still shows that the accuracy is much better than the Random Guess baseline. In addition, it is worth noting that using abstracts alone to predict the result is much faster than using the whole paper.

\subsection{Results of ChatGPT}
We performed experiments using the OpenAI API to assess ChatGPT's performance. Two types of inputs were tested: the full text of the paper and only the abstract.
Throughout the experiments, we encountered instances where our requests did not get a response from the OpenAI system due to their internal errors, as well as cases where ChatGPT's replies did not align with our prompts.
We filtered out these erroneous answers and focused solely on the appropriate correspondences.
The experimental results, presented in Table~\ref{tab:chatgpt}, indicate that the accuracy of ChatGPT as a reviewer is only comparable to random guessing, irrespective of the input type. Notably, ChatGPT demonstrated a tendency to predict "Accept" for all papers. We will further discuss these findings in the subsequent discussion section.

\begin{table}[tb]
\centering
\caption{\textbf{ChatGPT Results.} The "Answer" column represents the appropriate responses obtained. "Accept" and "Reject" columns indicate the count of Accept and Reject predictions generated by ChatGPT, respectively. }
\resizebox{.8\textwidth}{!}{
    \begin{tabular}{@{}lccccccc@{}}
    \toprule
    Input &  Answers &  \begin{tabular}[c]{@{}c@{}}Correct \\ Predictions\end{tabular} & Accuracy &  Accept &  Reject & \begin{tabular}[c]{@{}c@{}}Accuracy \\ of Big-4\end{tabular} & \begin{tabular}[c]{@{}c@{}}Accuracy\\ of arXiv\end{tabular} \\ \midrule
    Full text        &  1558 &   789 & 50.6\% & 1243 & 315 & 80.0\%  &  20.4\% \\
    Abstract         &  1675 &   929 & 55.5\% & 1442 & 233 & 90.2\%  &  18.3\% \\ 
    \bottomrule
    \end{tabular}
    }
    \vspace{-1.5em}
    \label{tab:chatgpt}
\end{table}

\section{Discussion}
\label{chap:discussion}

In this section, we discuss our experimental results, insights into predicting peer review outcomes, and the broader implications of employing AI methods in the peer-review process. 

\subsection{Discussion of Experimental Results}

\textbf{Algorithmic Fit}. As shown in Section~\ref{chap:experiment}, 12 of the 14 classification algorithms we have tested could obtain testing accuracy higher than $80\%$. The two ensemble models both obtain testing accuracy higher than $85\%$. 
In contrast, ChatGPT only achieves an accuracy of approximately $50\%$. 
These results demonstrate the basic effectiveness of Doc2Vec method. 
More specifically, SVM-based methods (with different kernel functions) all achieve high accuracy with a relatively reasonable amount of training and testing time. Notably, the SVM with a linear kernel demonstrates the highest accuracy and F1 score.

\noindent\textbf{Naive Rejecter}. We note that the accuracy results on the submissions subset are mostly between $80\%$ and $90\%$. Let us consider a naive classification algorithm, called the  ``rejecter'', which simply ``rejects'' every submitted paper in the ``Big-4'' conferences paper reviewing process. 
Due to the low overall acceptance rate of approximately 20\% in these conferences, the rejecter can achieve a similar accuracy of around 80\%.
However, this rejecter's high accuracy is irrelevant to our objective of exploring the predictability of the reviewing process in security conferences. Furthermore, the rejecter's performance does not undermine our experimental results and analysis.
Firstly, we intentionally constructed a balanced dataset to ensure that the classification algorithms have no prior knowledge of the acceptance rate, allowing them to evaluate each paper solely based on its content. Secondly, we thoroughly examined the prediction results (confusion matrix) of each classifier and found that they do not exhibit a bias towards accepting or rejecting papers. 
ChatGPT's tendency will be discussed separately.
Notably, most classifiers had a smaller number of False Negatives (FN) compared to False Positives (FP). However, in general, the FP and FN rates were similar.

\noindent\textbf{Impact of Novelty.}
A common concern about ML-supported scholarly review is that algorithms are not able to measure the novelty of a paper, or at least they are not as good as domain experts. This worry is legitimate %
because all existing automatic review models, including our model, are trained on historical data which inherently leads to model failure when they are dealing with unseen novel papers. 
In an attempt to test the ability of our method when dealing with the novelty aspect of papers, we design an experiment by excluding the recent years (2019--Now) of the ``Big-4'' security papers from our training dataset. Measuring the impact of novelty directly is intractable. The assumption and intuition behind our design is that as the security community advances year by year, the distribution gap between published papers and historical papers used for training becomes larger. For example, assuming that $\Gamma_{year}$ denotes the distribution of papers from a particular year %
and $\Gamma_{training}$ the distribution of all papers in the training set, then:  $\Gamma_{2022} - \Gamma_{training} > \Gamma_{2019} - \Gamma_{training}$. Papers with greater novelty presumably cultivate a larger gap compared to historical papers. Therefore, if our method is bad at predicting papers with larger delta, then it will not be good at predicting novel papers. 

We tested this hypothesis using our experiments and obtained the following mean accuracy values from each years' ``Big-4'' conferences: $90.2\%$ (2019), $90.3\%$ (2020), $88.2\%$ (2021), and $86.8\%$ (2022). 
The results demonstrate a (non-strict) declining tendency. Based on this declining tendency and general intuition, we presume that the novelty of a paper will weakly fertilize a negative impact on the classifier, making a novel paper more likely to be rejected, which is the opposite of what we want from the review process. Hence, the paper novelty aspect should be countered by a particular human-in-the-loop rating or general consideration. 

\noindent\textbf{Limitations}. 
Beyond weaknesses in dealing with novelty as discussed, we identified two limitations, both of which could make results favorable to higher accuracy. 
First, the papers we collect for training and testing are not from real ``Big-4'' submissions, which makes prediction easier than in the real scenario because the distribution of published ``Big-4'' papers is different from the distribution of selected arXiv preprint papers. Whereas in the real-world ``Big-4'' reviewing process, the submitted papers are more similar, making outcomes harder to predict. 
Second, our text normalization strategies, especially anonymization and publication information removal, are rather simple heuristics that may not be able to cover all cases. Therefore, we cannot rule out that some classifiers learn how to distinguish these corner cases in training.  

\noindent\textbf{ChatGPT Results}.
While ChatGPT is a highly advanced language model, it is primarily designed as a general-purpose chatbot and not specifically trained for reviewing papers. 
Several potential explanations can be considered to account for its poor performance:\\
1) \emph{Lack of domain-specific knowledge.} ChatGPT's training is based on a diverse range of internet text, limiting its knowledge in specific research domain necessary for proficient paper reviewing. Reviewing papers often requires expertise and understanding of the research landscape within a specific field or subject that ChatGPT may not possess to the same extent.\\
2) \emph{Inherent bias:} One potential factor contributing to the observed pattern of significantly more accept predictions than reject predictions by ChatGPT could be the presence of a bias in its response generation. ChatGPT is programmed to adopt a positive and polite attitude, which may influence its tendency to favor accept predictions over reject predictions.\\
3) \emph{Lack of in depth contextual understanding:} 
While ChatGPT can generate coherent responses based on the input it receives, it may not fully grasp the nuances and subtleties involved in security papers. Reviewing papers requires a deep understanding of the research methodology, experimental design, statistical analysis, 
which may be beyond the scope of ChatGPT's capabilities. \\
4) \emph{Forgetting:}
ChatGPT's working memory is notably constrained, leading to a limited capacity for retaining detailed information and contextual understanding within lengthy academic texts.\\
5) \emph{Hallucination:}
One notable limitation of ChatGPT is its tendency to generate outputs that may sound plausible but lack a genuine understanding of the input. This phenomenon, known as hallucination, poses a significant drawback to the reliability of ChatGPT's responses. We observed instances of hallucination when we noticed inconsistent predictions from ChatGPT for the same paper, with conflicting recommendations for acceptance and rejection.\\
6) \emph{Token limitation:}
ChatGPT 3.5 is constrained by a token limit of 4K, necessitating the segmentation of complete paper texts into smaller chunks. While researchers have proposed transformers with extended token capacities~\cite{bertsch2023unlimiformer}, ChatGPT has not yet incorporated these advancements.

In contrast to the proposed methods, ChatGPT possesses an advantage in its ability to provide explanations for the acceptance or rejection of papers. However, these explanations may also be subject to hallucination and lack a true understanding of the underlying reasons.

\subsection{AI and Peer Review}
Academic communities rely on peer-reviewing to decide on the acceptance of newly written papers and provide useful suggestions for improving research works.
In the journey of practicing it, researchers have also recognized the pros and cons of such systems~\cite{soneji2022sp,sun2021does,baxt1998reviews,smith2006peer}.
In this section, we would like to discuss the current review system, possible improvements to it, and how ML methods 
could be used to improve the review system.

\noindent \textbf{Peer Review --- An Imperfect Working System.} %
As Soneji~\etal concluded in their paper~\cite{soneji2022sp}, the current peer-review system is ``\textit{flawed, but we don't have a better system}''. 
The reviewers, just like any other human beings, could be affected by non-academic factors when they are making decisions, not to mention different reviewers could have totally different, but still well-justified judgments on the same paper. 
It is natural for researchers to be curious about the consistency in peer-review process. 
To the best of our knowledge, this kind of study is yet to be done in cyber-security communities. 
But there is a famous ``the NIPS experiment'' from Cortes and Lawrence in the ML community, who made 1/10th of the papers submitted to NIPS 2014 be passed through the review process twice independently for analyzing the review outcomes~\cite{theNIPSexperiment}. They found a consistency of the NIPS 2014 review process of just $25.9\%$, and only $43\%$ of papers accepted to the conference would be accepted again if the conference review process were repeated~\cite{tran2020open}. 
The result is so interesting that many researchers gave comments and participated in discussions~\cite{blogsontheNIPSexp}. 
Similar experiments and analysis have been conducted since~\cite{Hannah20How,shah2018design,tran2020open}, with similar results.

\noindent \textbf{Explorations toward a Better Peer-Review System.} %
Researchers are innovators and they do not stop exploring. The computer science community has made efforts to explore a better peer-review system. 
OpenReview\footnote{\url{https://openreview.net}} has appeared as one of the most noticeable new review systems in the last decade.
It was created to promote the open peer-review model in the computer science community, which means the submitted papers, reviewer's comments, authors' replies and other materials relating to the review process are open online to the public so that more researchers get access to the traditionally undisclosed review process. 
It seems that such open strategy does improve the peer-review system in terms of consistency~\cite{tran2020open} and also gives researchers the opportunity to analyze the review system for finding improvement measures. 

The ``Big-4'' security conferences have not experimented with an open peer-review model. Smaller security conferences occasionally have, an example being ACM WiSec 2016 that experimented making meta reviews of accepted papers publicly available  on their website.\footnote{\url{https://www.sigsac.org/wisec/WiSec2016/accepted-papers.html}} 
Although this is not a fully open peer-review paradigm, it can be considered as a step of the cyber-security community in trying to promote a peer-review system toward more openness. We would like to encourage the use of OpenReview in the cyber-security community to explore if we can together make the review process more open. An open peer-review process not only facilitates communication among researchers, but also allows more analysis on review system and better data for the researchers.

\noindent \textbf{Involve AI in Peer-Review Systems.} %
The advancements ML and AI have opened up possibilities for tackling complex real-world tasks with more intelligent algorithms. 
This progress prompts an intriguing and promising exploration of how ML can enhance the peer-review system, not only to alleviate the reviewers' workload but also to improve its overall effectiveness.
While it is evident from our experimental results and discussions that no ML or AI model can replace human reviewers' capabilities, there is still potential for ML methods to serve as supplementary or complementary components within the peer-review system.
Notably, researchers have been investigating ML novelty detection~\cite{pimentel2014review}, a task for detecting test data that are different from training data in some aspects. 
Particularly, Amplayo~\etal proposed a network-based approach to detect novelty of scholarly literature~\cite{amplayo2018network}. 

While the detection of novelty, contributions, and other merits may still be in an early stage of development, ML-based tools are already assisting human peer-review processes in various ways. 
For example, there are established approaches for detecting academic plagiarism~\cite{foltynek2019academic}.
Additionally, 
reviewers can leverage ML-powered tools like Grammarly and ChatGPT to help with their assessment of writing quality and clarity.
With the help of these existing tools, reviewers are encouraged to focus more on the paper's novelty, contribution, correctness, and prospect of a paper as they can allocate less time worrying about plagiarism and typos.
By offering positive incentives, we believe that AI tools play a beneficial role in enhancing the peer-review system.

\section{Conclusion}
\label{chap:conclusions}
To study the predictability of  peer reviews for security papers, we tested the reviewing capabilities of a Doc2Vec-based method and a ChatGPT-based method on top-tier conferences on computer security and privacy. 
The results demonstrate that Doc2Vec-based method achieves approximately 90\% prediction accuracy, while ChatGPT achieves only 50\% accuracy.
While our method exhibits reasonable accuracy in predicting reviewing outcomes for security research papers, there is a noticeable error rate. In the meantime, utilizing AGI models like ChatGPT for academic review still requires substantial advancements.
We conclude that our proposed method could predict the reviewing outcomes for security research papers with reasonable accuracy, but the error rate is non-negligible. 
Despite acknowledging the limitations of our method, we explore the AI performance in cybersecurity peer-review task and discussed our findings with regards to the peer-review system. In conclusion, the peer review system has successfully adapted to numerous challenges over the past decades. We hope our work encourages further research to address the evolving challenges faced by peer review in the AI era.

\balance
\bibliographystyle{splncs04} %
\bibliography{secreview}

\appendix

\section{Common Peer-Review Process}
\label{app:review_process}
Peer review is a crucial process of computer science conferences, aiming to assess papers submitted to corresponding venues by other experts or peers. As an indispensable part of the scholarly research cycle, peer review plays a crucial role in %
publishing and disseminating state-of-the-art research and results.%

\begin{figure*}[h]
    \centering
    \includegraphics[width=1\linewidth]{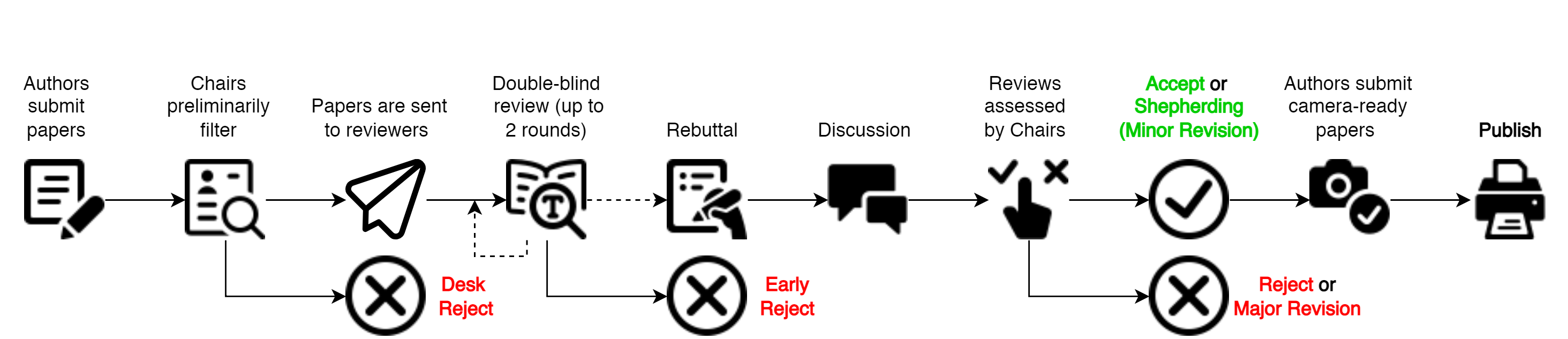}
    \caption{The paradigm of the peer-review process for ``top-tier'' security conferences. This diagram includes all steps of the post-submission life cycle of a paper.
    Dashed lines indicates optional procedures. Bold and color fonts indicate decisions. Different conferences vary in the details, but in general, they all conform to a similar paradigm. For example, ``Accept on Shepherd Approval'' is similar to ``Conditional Accept or Minor Revision''. ``Early Reject'' is essentially similar to ``Reject'', but those papers do not reach the second round of review.
    }
\label{fig:peerreivew}
\end{figure*}

Many security conferences have turned to a more journal-style peer-review process in recent years (the entire review process is depicted in Figure~\ref{fig:peerreivew}).
The final decision now includes not only accept and reject, but also revision.
Additionally, the number of review cycles per year increased: %
CCS %
and NDSS %
now have two review cycles per year; USENIX Security has three rolling review cycles, %
the same number that IEEE S\&P %
has recalibrated to from an earlier per-month review cycle with 12 submission deadlines per year. The review is double-blind across all the review stages, meaning the identities of authors and reviewers are hidden from each other. The review process %
happens in multiple review rounds; papers that make the first round get additional reviewer assignments.

\begin{table}[htb]
	\centering
 	\caption{Review outcomes for the ``Big-4'' conferences.}
            \begin{tabular}{@{}lc@{}}
    		\toprule
                \centering
    		\textbf{Conference} & \textbf{Final Review Outcomes}\\ \midrule
    		ACM CCS             & Accept/Minor Revision/Reject \\ \midrule
    		IEEE S\&P           & Accept/Conditional Accept/Major Revision/Reject           \\ \midrule
    		ISOC NDSS           & Accept/Minor Revision/Major Revision/Reject \\ \midrule
    		USENIX Security     & 
                \begin{tabular}[c]{@{}c@{}}Accept/Accept on Shepherd Approval/\\ Accept Conditional on Major Revision/Reject\end{tabular} 
    		\\ \bottomrule
            \end{tabular}
\label{tab:bigfourresult}
\end{table}

An overview of the peer-review process for top computer conferences on security and privacy is shown in Fig~\ref{fig:peerreivew}.

\section{Prompt of ChatGPT Experiment}
\label{app:prompt}
We use the following prompt to let ChatGPT act as a reviewer:
\begin{quote}
You are an experienced and fair reviewer from top cybersecurity conferences (NDSS, IEEE S\&P, CCS and USENIX Security). I will give you a paper for you to read and review. Due to the token limitation, I will split the paper content into some chunks and I will let you read the entire paper chunk by chunk. Please only reply with "OK" if the text does not contain "<|end\_of\_paper|>". Once you receive it, please merge the previous messages together into a full paper for you to review. I want you to decide whether this paper should be accepted or not. You must first tell me your decision with "Accept" or "Reject", and then explain your reasons in concise language.
\end{quote}

\section{Composition of Proceedings Subset}
Table~\ref{tab:cpd} contains the information of the proceedings we collect.
\begin{table*}[h]
\centering
\caption{\textbf{Proceedings subset.} All papers accepted and published by the corresponding conferences.}
\resizebox{1.\textwidth}{!}{
\begin{tabular}{@{}llr|llr|llr@{}}
\toprule
\textbf{Venue} & \textbf{Area} &\textbf{\#Samples} & \textbf{Venue} & \textbf{Area} & \textbf{\#Samples} & \textbf{Venue} & \textbf{Area} & \textbf{\#Samples} \\ \midrule
CCS 2012       & Security              & 81   & ICCV 2013    & Computer Vision       & 454 &  SP 2013              & Security & 38  \\   
CCS 2013       & Security              & 105  & ICCV 2015    & Computer Vision       & 526 &  SP 2014              & Security & 44  \\   
CCS 2014       & Security              & 114  & ICCV 2017    & Computer Vision       & 621 &  SP 2015              & Security & 55  \\   
CCS 2015       & Security              & 128  & ISCA 2018    & Computer Architecture & 64  &  SP 2016              & Security & 55  \\   
CCS 2016       & Security              & 137  & MobiCom 2017 & Networking            & 35  &  SP 2017              & Security & 60  \\   
CCS 2017       & Security              & 151  & MobiCom 2018 & Networking            & 42  &  SP 2018              & Security & 63  \\   
CCS 2018       & Security              & 134  & NDSS 2013    & Security              & 47  &  SP 2019              & Security & 82  \\   
CCS 2019       & Security              & 149  & NDSS 2014    & Security              & 55  &  SP 2020              & Security & 104 \\   
CCS 2020       & Security              & 121  & NDSS 2015    & Security              & 50  &  SP 2021              & Security & 115 \\   
CCS 2021       & Security              & 196  & NDSS 2016    & Security              & 60  &  SP 2022              & Security & 147 \\   
Crypto 2018    & Security              & 79   & NDSS 2017    & Security              & 68  &  STOC 2018            & Theory   & 111 \\   
Crypto 2019    & Security              & 81   & NDSS 2018    & Security              & 71  &  USENIX Security 2013 & Security & 44  \\
CVPR 2013      & Computer Vision       & 471  & NDSS 2019    & Security              & 89  &  USENIX Security 2014 & Security & 67  \\
CVPR 2014      & Computer Vision       & 540  & NDSS 2020    & Security              & 88  &  USENIX Security 2015 & Security & 67  \\
CVPR 2015      & Computer Vision       & 602  & NDSS 2021    & Security              & 90  &  USENIX Security 2016 & Security & 71  \\
CVPR 2016      & Computer Vision       & 643  & NDSS 2022    & Security              & 83  &  USENIX Security 2017 & Security & 85  \\
CVPR 2017      & Computer Vision       & 783  & PPoPP 2018   & Parallel Programming  & 28  &  USENIX Security 2018 & Security & 100 \\
CVPR 2018      & Computer Vision       & 979  & SIGCOMM 2015 & Networking            & 43  &  USENIX Security 2019 & Security & 112 \\
EuroCrypt 2018 & Security              & 69   & SIGCOMM 2016 & Networking            & 39  &  USENIX Security 2020 & Security & 156 \\
EuroCrypt 2019 & Security              & 76   & SIGCOMM 2017 & Networking            & 38  &  USENIX Security 2021 & Security & 246 \\
FOCS 2018      & Theory                & 86   & SIGCOMM 2018 & Networking            & 40  &  USENIX Security 2022 & Security & 256 \\
HPCA 2019      & Computer Architecture & 37   & SIGKDD 2015  & Data Mining           & 48  &  \textbf{Total} & & \textbf{10,519}  \\   %

\bottomrule
\end{tabular}
}
\label{tab:cpd}
\end{table*}

\end{document}